\documentclass[twocolumn,twocolappendix]{aastex631}
\usepackage{graphicx}
\usepackage{grffile}
\usepackage{dcolumn}
\usepackage{xcolor}
\usepackage{color}
\usepackage{bm}
\usepackage{amssymb, amsmath}
\usepackage{soul}
\usepackage{times}
\usepackage{cancel}

\graphicspath{{./}{figures/}}

\begin{document}
	
	\preprint{XXX}
	
	\title{On the Effect of Large Scale Structures and Turbulence on Solar Eruptive Events: The Cross-Scale Challenge for Space Weather}

    \author[0000-0003-3823-127X]{Alessandro Ippolito}
    \affiliation{Istituto Nazionale di Geofisica e Vulcanologia (INGV), Rome, Italy}
	
	\author[0000-0003-3739-3170]{Giuseppe Prete}
	\affiliation{Dipartimento di Fisica, Universit\`a della Calabria, Arcavacata di Rende, 87036, Italy}
	
	\author[0000-0001-6344-6956]{Nicolas Wijsen}
	\affiliation{Centre for Mathematical Plasma Astrophysics, Dept. of Mathematics, KU Leuven, Celestijnenlaan 200B, 3001 Leuven, Belgium}
	
    \author[0000-0002-9207-2647]{Gaetano Zimbardo}
	\affiliation{Dipartimento di Fisica, Universit\`a della Calabria, Arcavacata di Rende, 87036, Italy}

        \author[0000-0002-4269-056X]{Anwesha Maharana}
	\affiliation{Centre for Mathematical Plasma Astrophysics, Dept. of Mathematics, KU Leuven, Celestijnenlaan 200B, 3001 Leuven, Belgium}
	
	\author[0000-0002-1743-0651]{Stefaan Poedts}
	\affiliation{Centre for Mathematical Plasma Astrophysics, Dept. of Mathematics, KU Leuven, Celestijnenlaan 200B, 3001 Leuven, Belgium}
    \affiliation{Institute of Physics, University of Maria Curie-Skłodowska, ul.\ Radziszewskiego 10, 20-031 Lublin, Poland}
	
	\author[0000-0001-8184-2151]{Sergio Servidio}
	\affiliation{Dipartimento di Fisica, Universit\`a della Calabria, Arcavacata di Rende, 87036, Italy}
	
	\begin{abstract}
Coronal Mass Ejections (CMEs) are among the most powerful drivers of space weather, yet their prediction remains elusive. A fundamental obstacle is the problem's multiscale nature: large-scale magnetohydrodynamic models do not resolve the turbulent fluctuations that govern particle transport and magnetic connectivity. We present a new model that combines a MICroscopic diffusion approach of turbulence with MACroscopic EUHFORIA simulations (MICMAC) to address the cross-scale challenge of space weather. By incorporating turbulence properties into such small-scale Monte Carlo simulations, we describe an extreme CME event, tracing both magnetic field lines and 100 MeV protons from the CME-driven shock to 1 AU. We find that turbulence dramatically broadens and distorts the magnetic connection between the CME nose and Earth, producing footpoint distributions that are highly non-Gaussian, anisotropic, and patchy. MICMAC suggests that the distribution's enstrophy (non-Gaussianity) grows as the CME approaches Earth, signaling anomalous diffusive behavior. A simple shear-layer toy model reproduces the observed in-plane anisotropy, suggesting that local current sheet geometry imprints a persistent memory on the turbulent field. Our results demonstrate that cross-scale coupling between the CME's large-scale structure and ambient turbulence must be accounted for in space weather models. We discuss implications for SEP forecasting and interpreting multi-spacecraft observations.
	\end{abstract}
	
	\keywords{Space plasmas -- turbulence -- space weather -- coronal mass ejections -- solar energetic particles}
	
\section{Introduction}
\label{sec:intro}	
Space weather forecasting faces a fundamental obstacle: it involves a vast range of spatial and temporal scales, from coronal structures to kinetic dissipation lengths. Explosive events like Coronal Mass Ejections (CMEs) expand through the solar wind, encountering a ubiquitous and unavoidable complication---turbulence. In this cross-scale picture, large-scale fields, though often well constrained by solar observations, are influenced by unresolved scales in global magnetohydrodynamic (MHD) models \citep{Pomoell_Poedts_2018}. Computational power, however impressive, is never sufficient; finite resolution inevitably introduces mismatches between simulations and reality. Our aim is to mitigate this gap by building a bridge between large-scale mesoscopic models of extreme events and the theory of plasma turbulence.
	
For decades, the space plasma community has operated along two apparently disconnected tracks. First, turbulence has been studied as a paradigmatic challenge, with elegant theories describing a ubiquitous state of matter spanning from ordinary fluids to distant galaxies. Second, shocks have been investigated as quasi-steady barriers in front of planets—such as the Earth's bow shock—or as the leading edge of impulsive, energetic events like CMEs, which drive Solar Energetic Particles (SEPs) that endanger spacecraft, astronauts, and telecommunications. 
These two ingredients are, in reality, a single intertwined system. Turbulence creates the magnetic landscape of current sheets, waves, vortices, and flux ropes that CME shocks compress and distort \citep{Servidio2009, Servidio2010, Matthaeus2015, Trotta2022, Trotta2023}. Shocks, in turn, inject free energy, amplifying turbulent fluctuations and accelerating particles \citep{Effenberger2025}. This Letter brings these aspects together in a unified framework.
	
The key challenge is to understand extreme events such as interplanetary shocks in the presence of turbulence and to determine how small-scale processes feed back on field-line diffusion and, hence, particle transport. The number of observations of such interactions is remarkable. While CME propagation is routinely modeled with empirical or large-scale MHD codes, both approaches neglect the multi-scale turbulent nature of the interplanetary medium and CMEs \citep{Pomoell_Poedts_2018}. Recent works demonstrate that turbulent structures survive shock crossings carrying magnetic helicity downstream \citep{Trotta2022}, and that pre-existing turbulence dramatically modifies shock properties and particle transport \citep{Giacalone1999, Laitinen2013, Laitinen2016, joubert2026turbulent}. Moreover, \citet{Wijsen2019} showed that complex solar wind configurations, such as Corotating Interaction Regions (CIRs), can significantly modify particle transport, with compression regions acting as magnetic mirrors and enhancing cross-field diffusion even with small perpendicular mean free paths.
	
In this work, we present the MICroscopic diffusion model of turbulence, combined with MACroscopic EUHFORIA simulations (MICMAC), to address the cross-scale challenge in space weather. MICMAC complements EUHFORIA with a subgrid turbulence model based on non-quasilinear Monte Carlo transport \citep{Veltri1998, Zimb2000, Pommois2001, Ippolito2005}. We apply this framework to the November 3, 2021 CME \citep{Soni2024}, tracing magnetic field lines and 100 MeV protons from the CME-driven shock to 1~AU. We show that cross-scale coupling dramatically alters magnetic connectivity and particle propagation, with implications for SEP forecasting.
	
\begin{figure}
\centering
\includegraphics[width=0.99\columnwidth]{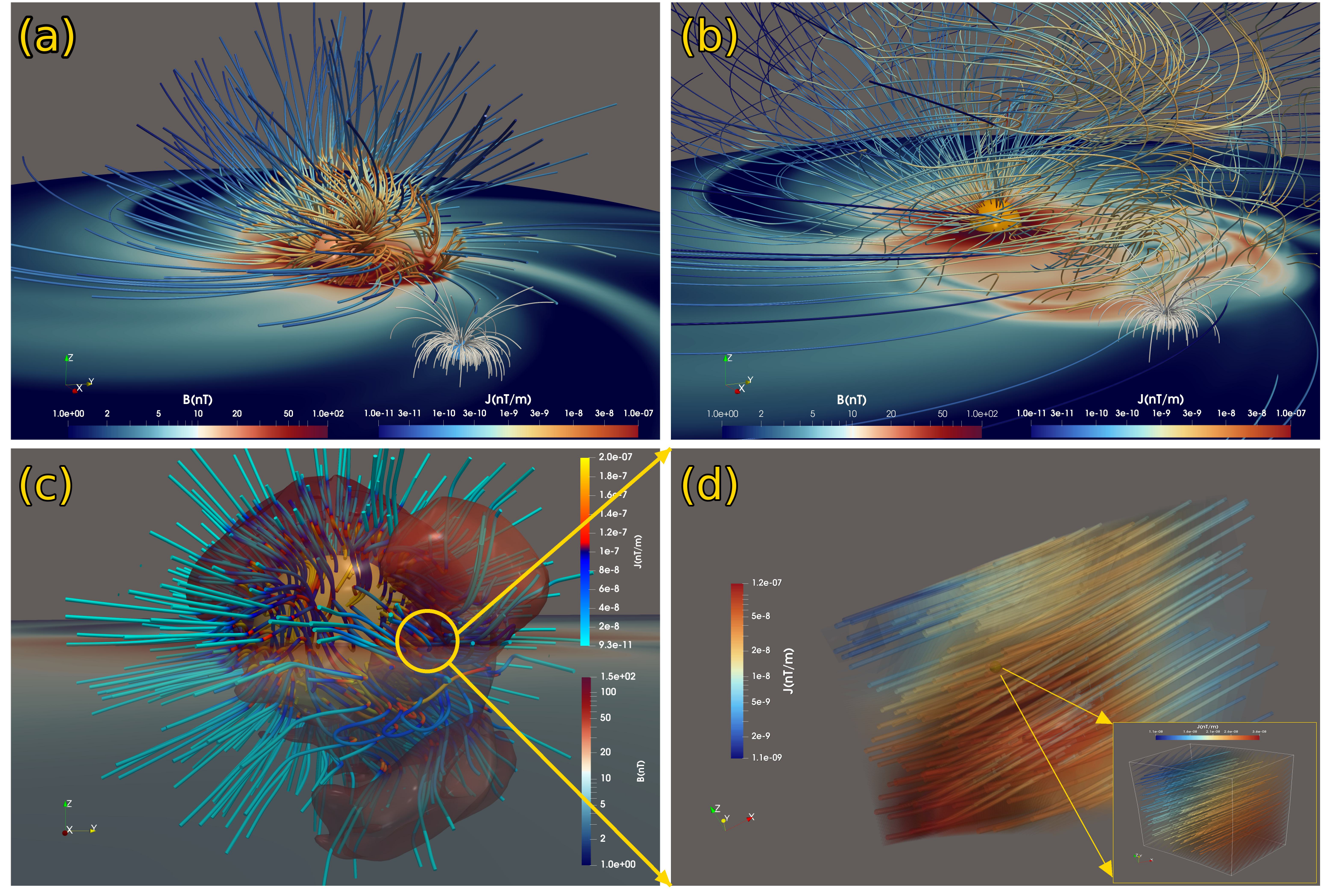}
\caption{\textit{Upper row:} Evolution of an extreme CME with field lines and currents (shaded contours),   at t=19:59:40 of 02/21 (a) and t=01:59:39 of 04/21 (b). Lines (white) around Earth (blue sphere) is representative of Earth's magnetic field. Panel (c) shows the complex pattern of the currents (isocontours) at possible injection regions of solar energetic particles, while the zoom (d) suggests a strongly radial, sheared magnetic field (high current) in front of the emerging structure. In this sub-volume, we start the tracing.}
\label{fig1:cme}
\end{figure}

\section{Methods}
\label{sec:methods}	
{\it The macroscopic model.} We investigate the transport of interplanetary magnetic field lines and particles during the CME events of November 3, 2021. We focus on the event analyzed in \citet{Soni2024}, which involves the merging of multiple CMEs. This event is particularly useful for forecasting geomagnetic disturbances, as Solar Orbiter (SolO) and ACE were aligned along the same Sun-Earth line during the CME passage. To reproduce the evolution and merging of these three CMEs, we employ the Spheromak model \citep{Kataoka2009JGRA, verbeke2019A&A} within the EUHFORIA framework. This model allows us to prescribe an internal flux-rope structure for each spherical CME. We adopt the parameters from \citet{Soni2024}.

Figure~\ref{fig1:cme} shows the evolution of a realistic extreme CME at two different times. The CME hits the Earth, but not with the CME’s nose; rather, the interaction occurs in the region below the nose. 
The CME hits Earth, but not with its nose; instead, the interaction occurs below the nose, indicating a moderate-risk case in which the mesoscopic field does not directly impact our planet.
The question then becomes: how does turbulence affect predictions and the dynamics of field lines and particles in an event where the space weather risk is only moderate? Does turbulence influence the predictability and dispersion of high-energy particles?

\begin{figure*}
\centering
\includegraphics[width=2.10\columnwidth]{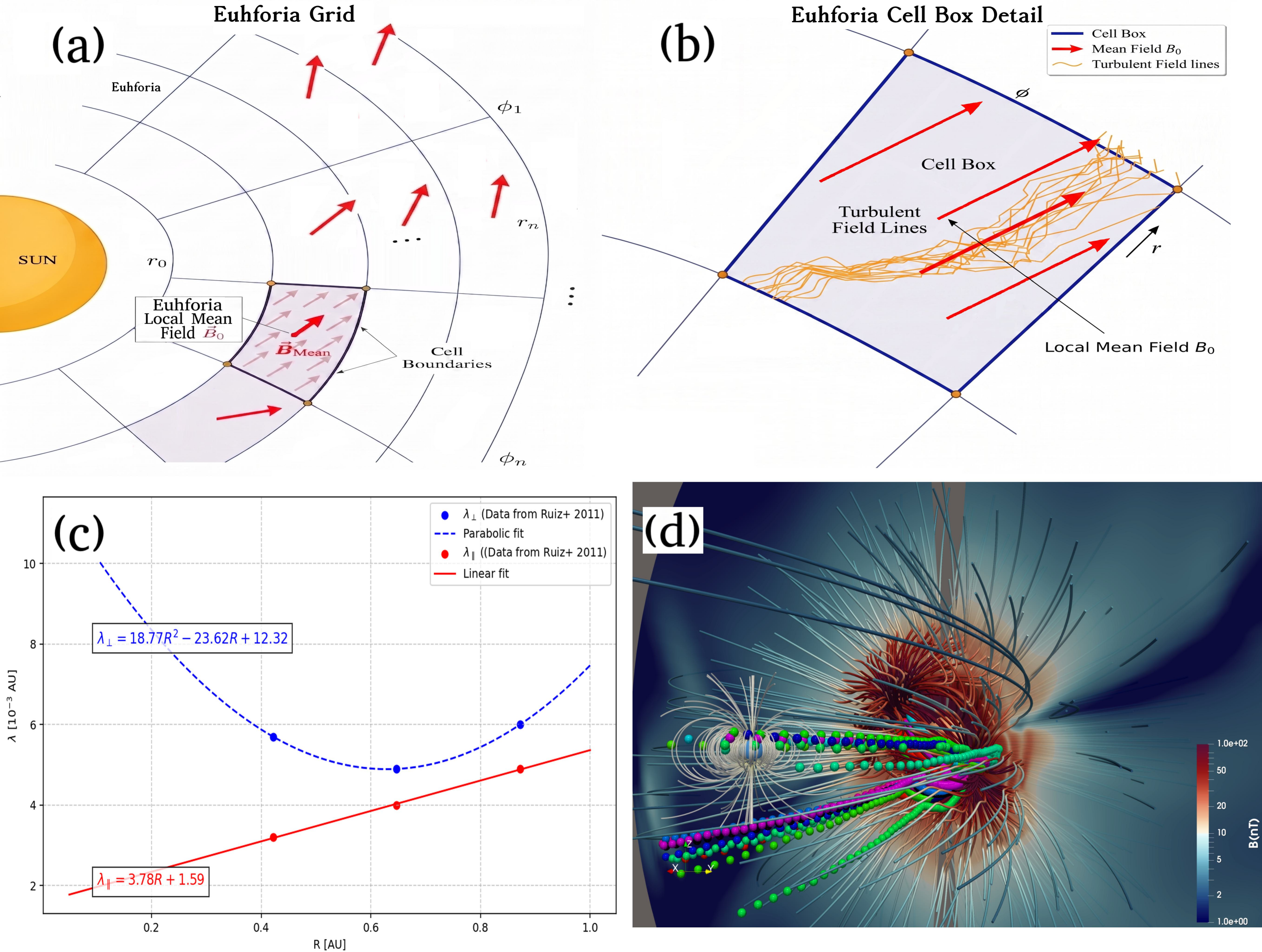} 
\caption{(a) Cartoon of the EUHFORIA grid, representing the cells (black open circles) and their boundaries, with the ambient magnetic field ${\bf B}_0$ (red thick arrows), (b) Zoom in of the EUHFORIA cell (shaded), with the turbulent field lines superimposed. The lines propagate through the Fokker-Planck Eq.s (\ref{eq:fieldline})-(\ref{eq:di}) around the ambient mesoscopic field $B_0$. (c) Recipes of the correlation lengths as a function of the solar distance, from \citet{Ruiz2011}. (d) Final results of the combination of mesoscopic and microscopic turbulence modeling, looking from Earth. Particles computed in the turbulent models are represented as thick spheres.}
\label{fig2:model}
\end{figure*}

{\it The microscopic model.} To reconstruct the microscopic evolution of field lines from the solar corona to the surrounding heliosphere, we employ a Monte Carlo (MC) code \citep{Pommois2001, Ippolito2005, Ippolito2022, Ippolito2025}. Since the infinitesimal displacement $d\mathbf{r}$ is parallel to $\mathbf{B(r)}$, such that $d\mathbf{r} = \mathbf{B}d\gamma$, treating $d\gamma$ as $d\xi/B_0$, with $B_0$ a constant, we obtain:
\begin{equation}
	\frac{d\mathbf{r}}{d\xi} = \frac{\mathbf{B(r)}}{B_0},
	\label{eq:fieldline}
\end{equation}	
where the total magnetic field $\mathbf{B}(\mathbf{r})$ is composed of the mean field $\mathbf{B}_0 = \mathbf{B}_{0z}(\mathbf{r})$ and the fluctuating component $\delta \mathbf{B}(\mathbf{r})$.
	
Equation~\eqref{eq:fieldline} is integrated within a local Cartesian reference frame: the $z$-axis is aligned with the local mean interplanetary magnetic field $\mathbf{B}_0$; the $x$-axis is perpendicular to the plane containing the radial direction and $\mathbf{B}_0$; the $y$-axis completes the right-handed coordinate system. To account for magnetic field-line diffusion under anisotropic turbulence—as typically found in the solar wind \citep{Goldstein1995}— we implement a three-dimensional model of magnetic turbulence. The MC simulation adds a random force term, scaled by the diffusion coefficient, to the field-line equations. The fluctuating magnetic field component is modeled as:
	\begin{equation}
		\frac{\delta B_i(r)}{B_0(r)} = \eta_i(\xi)A_i(r),
	\end{equation}
where $i = x, y$, $\eta_x(\xi)$ and $\eta_y(\xi)$ are uncorrelated stochastic functions, and $A_i(r) = \sqrt{6 D_i(r)}$ denotes the amplitudes of the random forces \citep{Veltri90, Veltri1998, Pommois2001, Ippolito2005}. The diffusion coefficients in the local Cartesian frame are defined as:
	\begin{equation}
		D_i = \mathcal{D}\left(\frac{\delta B}{B_0} \frac{l_z}{l_x}\right)^\mu \frac{l_x^2}{l_z} \left(\frac{l_i}{l_x}\right)^\nu,
		\label{eq:di}
	\end{equation}
where $\mathcal{D} = 0.028$, $\mu = 1.51$, and $\nu = 0.67$ \citep{Zimb2000,Pommois2001}. The correlation lengths $l_x, l_y, l_z$ quantify the turbulence anisotropy. Typical fluctuation levels in the solar wind at 1~AU range between $\delta B/B_0 \simeq 0.5-1$ \citep{ZanK1998}, corresponding to anisotropy ratios of $l_x/l_y = 1 - 10$ and $l_z/l_y = 1 - 10$. For each cell within the EUHFORIA grid, we determine $\lambda_\parallel$ and $\lambda_\perp$ by fitting the data of \citealt{Ruiz2011}, as illustrated in Figure~\ref{fig2:model}-(c). In our model $\lambda_\parallel = \ell_z$ and $\lambda_\perp = \ell_x$, since we choose $\ell_x=\ell_y$.

\subsection{The MICMAC Model}
\label{sec:model}
To investigate the transport of interplanetary magnetic field (IMF) lines during the CME event, we utilize time-dependent plasma parameters ($V_{sw}$ and $\mathbf{B}$) extracted from EUHFORIA simulation grids, tracking the CME propagation from 0.1 to 0.95~AU. We trace 1000 individual field lines, incorporating a random-walk component to account for the stochastic effects of solar-wind turbulence. In addition, we perform MC simulations to calculate the trajectories of 100~MeV protons, accounting for the magnetic-field-line random walk induced by solar-wind turbulence. The heliospheric magnetic field lines are modeled using such ``mesoscale'' field $B_0$ generated by the EUHFORIA simulation \citep{Pomoell_Poedts_2018}, as shown in a typical CME eruption in Fig.~\ref{fig1:cme} (a)-(b).

We identify spatial regions where the CME-driven shock generates maximum current density $ J$ (at heliocentric distances of 0.4 and 0.7~AU). Using a Boris integration method, we track proton trajectories originating from these $J_{\max}$ zones, employing the diffusion coefficients defined above to characterize the random walk of the particles' guiding centers. Such high-current regions are important candidates for magnetic reconnection and, hence, particle-seeding mechanisms for gradual SEPs. Moreover, MC simulations provide an easily implementable, fast, and adaptable technique for real-time forecasting models.

The most important regions that are sources of CME-accelerated SEPs are at the shock, where reconnection also occurs, as can be seen from Figure~\ref{fig1:cme}-(c), where we show the complex pattern of the magnetic field together with the network of high-current layers that surround the CME. Panel (d) shows a zoom of the ambient field. In such a high-current region, the magnetic field is almost perfectly radial, with a transverse shear: an in-plane component generates a current layer that permeates the starting point of the field lines that will most likely guide high-energy particles. In this complex network of structures and inhomogeneities, we begin our tracing.

The strategy combines the mesoscopic EUHFORIA code with subgrid turbulence modeling. The philosophy is straightforward: the large-scale field provides ambient guidance to the turbulence model. Each element of the EUHFORIA volume has an associated value of the magnetic field $\mathbf{B}_0(r, \theta, \phi)$, which changes from cell to cell at a given time. In full consistency with the finite-volume numerical method, we assign a constant field value within each cell. The physical size of these cells is comparable to (slightly smaller than) the ambient turbulence correlation length, which is ideal for matching the models. Figure~\ref{fig2:model}-(a) shows a cartoon of the EUHFORIA grid, with cells represented by black open circles and boundaries. The ambient magnetic field $\mathbf{B}_0$ (thick arrows) varies from cell to cell. Panel (b) shows a zoom of a single EUHFORIA cell (shaded), with turbulent field lines superimposed. The lines propagate according to the diffusion equation around the ambient mesoscopic field $B_0$, exhibiting the characteristic meandering of a random walk. The lines continue their journey, feeling variations in both the large-scale fields and the turbulence. The latter incorporates observational ingredients such as the anisotropy of the correlation scales and the fluctuation level $\delta B/B_0$. We use the radial distribution of correlation scales as described by \citealt{Ruiz2011}, employing a polynomial fit of the data as shown in Figure~\ref{fig2:model}-(c), which we use to prescribe the turbulence properties as a function of heliocentric distance.
	
\begin{figure*}
\centering
\includegraphics[width=1.20\columnwidth]{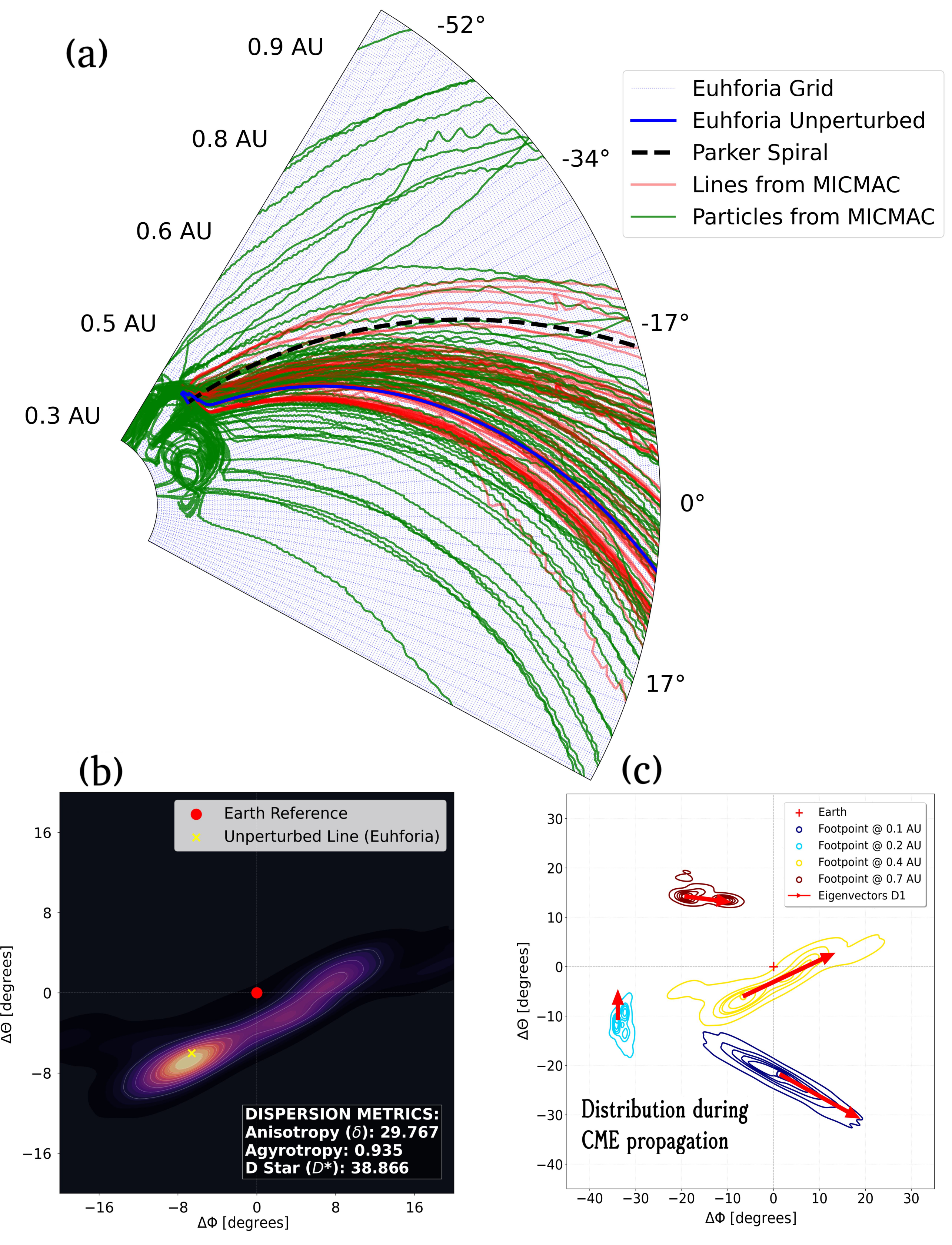}
\caption{(a) Comparison between the Parker spiral (black dashed), the unperturbed mesoscopic EUHFORIA  field (thick red) and the combination of large-scale inhomogeneous dynamic field with subgrid turbulence modelling in MICMAC (orange). In the MICMAC case, particles experience dramatic scattering, clustering, and dispersion. (b) Field line distribution at 1 AU, in the azimuthal-polar plane $f_{1AU}(\theta, \phi)$. The distribution, driven by a complex large-scale magnetic skeleton, is highly anomalous at these distances, exhibiting angular anisotropy and agyrotropy. The plane perpendicular to the mean field, which does not exactly intercept the theta-phi plane, shows analogous results. (c) Distributions of the field lines $f_{1AU}(\theta, \phi)$ for different starting points, following the CME propagation. Eigenvectors of the main in-plane anisotropy are shown with red arrows. During propagation, the anisotropy varies, the distribution tilts, and, moreover, for distances close to Earth, the distributions become more irregular and patchy, suggesting an immature diffusion that feels the coherency of the large-scale inhomogeneous CME-perturbed field.}
	\label{fig3:allcases}
\end{figure*}

We integrate the MC method into the subgrid by using the ambient magnetic field as the macroscopic constant. When a field line jumps to another cell, it feels the new ambient field, and the motion continues as a diffusive motion superimposed on the large-scale coherent field. The integration steps are chosen to be sufficiently small so that the field line visits only one cell at a time and remains long enough in a region to adequately introduce the effect of turbulence.  
The EUHFORIA background field satisfies the discrete solenoidal constraint. The stochastic component does not represent an additional magnetic-field realization, but statistically models field-line spreading around this background.
For radial excursions, every time the field line enters a new cell, it can be viewed as a separate Monte Carlo process with the new ambient values of the macroscopic fields. 
    Note that our model shares some similarities with \citet{joubert2026turbulent}, where they apply a closely related stochastic approach to a Parker field, starting from a convection–diffusion equation for the field-line density and deriving the corresponding SDEs. Importantly, their stochastic trajectories are described, as in our case, as pseudo-field lines: independent realizations are used to describe field line spreading.
	
Over this reconstructed field, we integrate particle trajectories at 100~MeV. Over large distances, MICMAC yields a natural spreading of protons, as illustrated in Figure~\ref{fig2:model}-(d), which shows the result of the combination of mesoscopic and microscopic turbulence modeling, looking from Earth. Particles (represented as thick spheres) have a tendency to cluster, reflecting the coherency of the large-scale field \citep{Sonsrettee2024}.

In Figure~\ref{fig3:allcases} we compare three scenarios: the classical Parker spiral, the unperturbed EUHFORIA field, and the combination of the large-scale inhomogeneous dynamic field with subgrid turbulence modeling as implemented in MICMAC. The difference between the unperturbed EUHFORIA field and the Parker prediction — due to the CME — is substantial. When turbulence is added, the spread of field lines becomes dramatic: bunches of field lines decouple from the main direction of $\mathbf{B}_0(\mathbf{r})$ and meander into other regions, also far from the target point of the central line. We also superimpose test particles (100~MeV protons) integrated over the EUHFORIA$+$turbulence field. The particle puffs seem to depend on the magnetic structure at the seeding region of the field lines (green dashed lines). The statistical behavior of field lines at 1~AU reveals a highly distorted probability distribution $f(\theta,\phi)$ over the sky map (Figure~\ref{fig3:allcases}-(b)). The distribution shows two main anomalies: a strong anisotropy and a multi-peaked structure. At this time, the field lines do not hit Earth. We compute the anisotropy and agyrotropy in the angular plane, finding values that suggest a strong influence of the large-scale EUHFORIA field over the turbulence diffusion \citep{Ruffolo2006}. Only by combining these cross-scale effects within MICMAC can one observe such anomalies, which have implications for observations and space weather.
	
We apply the model to different simulation times as the CME moves away from the Sun (Figure~\ref{fig3:allcases}-(c)). The distributions always show anisotropies in the plane with varying intensity. We report the dominant in-plane eigenvector, computed from $f(\theta,\phi)$, showing that the direction of the maximum eigenvalue varies with the CME position. Moreover, for times when the CME is closer to Earth, the distribution is more distorted and shows more subscale features.
	
\begin{figure*}
\centering
\includegraphics[width=1.50\columnwidth]{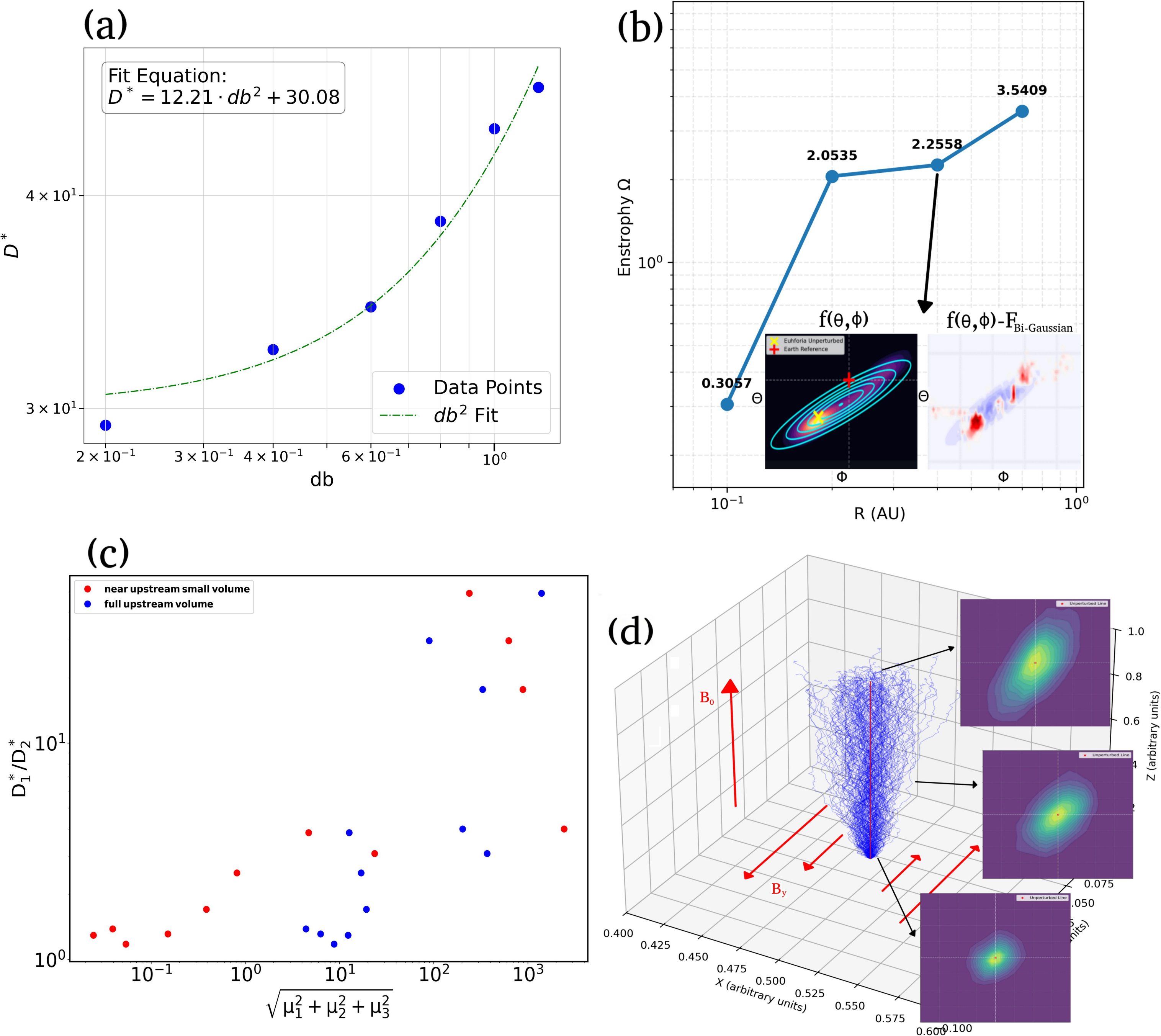}
\caption{(a) Influence of the diffusion coefficient on the level of magnetic fluctuation, showing agreement with the classical expectation for inertial range turbulent diffusion. (b) Enstrophy, defined in the text, showing that the distribution is highly non-Maxwellian for field lines starting from the CME nose when it is closer to Earth. The inset shows an example of the distribution and the deviation from bi-Maxwellianity, suggesting anomalous patchy diffusion driven by large-scale patterns. (c) Dependency of the transverse anisotropy on the minimum variance of the magnetic field at the seeding point (red) and at the full volume visited by the lines (blue). (d) Toy model for a simple current layer, reproducing the anisotropic behavior of the field lines.}
\label{fig4:diffproperties}
\end{figure*}

\subsection{Patchy Diffusion and Anisotropy}	
In order to comprehend the role of turbulence on our cross-scale diffusive model, we compute the trace of the variance tensor, 
\begin{equation}
D^*_{ij} = \int \int \left( \alpha_i-\langle\alpha_i\rangle\right)\left( \alpha_j-\langle\alpha_j\rangle\right)f(\theta,\phi) d^2 \alpha, 
		\label{eq:varang}
\end{equation}
where $(\alpha_1, \alpha_2)=(\theta, \phi)$, $\langle \alpha_i \rangle$ are the central moments of the distribution (average angles of the distribution in Figure~\ref{fig3:allcases} (b)) and $D^* = 1/2 Tr\{D^*_{i, j}\}$ is the width of the distribution. This quantity, in accordance with classical diffusive problems, is proportional to the magnetic field line diffusion. We observe that this parameter depends on the small-scale turbulence problem: larger relative fluctuations lead to broader distributions of field lines in space. This indirect measure of the diffusion coefficient depends on the level of turbulence, as shown in Figure~\ref{fig4:diffproperties}-(a), with a typical square-root dependency expected from classical theories of turbulent diffusion \citep{Matthaeus1995, Pommois2001}. This simple symmetric tensor admits two eigenvalues ($D^*_1, D^*_2$) and associated eigenvectors ($\hat{e}_1, \hat{e}_2$), revealing the topological properties of diffusion. The main eigenvectors $\hat{e}_1$ (along the broader diffusion direction) are shown in Figure~\ref{fig3:allcases}-(c).

To characterize the anomalies and substructures of the field line distributions — which might indicate coherency in the large-scale field lines — we compute the enstrophy of the distribution as 
\begin{equation}
\Omega = \frac{\int \left[ f - F_{BG} \right]^2 d\beta_1 d\beta_2}{\int \left[ F_{BG} \right]^2 d\beta_1 d\beta_2},
\label{eq:omega}
\end{equation}
where $\beta_i$ are the angular coordinates along the eigenvectors $\hat{e}_i$ of the variance tensor in Eq.~(\ref{eq:varang}), and 
$
F_{BG}(\beta_1,\beta_2) \sim e^{ -\frac{\beta_1^2}{2D^*_1} - \frac{\beta_2^2}{2D^*_2} }
$
is the bi-Gaussian, centered at the the average angles $\langle \alpha_i \rangle$, and built with the above eigensystem. This global quantity, the second Casimir invariant, quantifies the distribution's departure from equilibrium (in our case, a Gaussian diffusive state) \citep{Greco2012b}. Inhomogeneous kinetic effects, such as temperature anisotropies and non-Maxwellian distortions, concentrate near coherent structures in turbulent plasmas \citep{Greco2012b}. Figure~\ref{fig4:diffproperties}-(b) shows that when the CME is closer to Earth, the distribution is more modulated and gives rise to more anomalous patterns. The inset shows the distribution at a given time and its deviation from a bi-Gaussian. Such patchy transport regimes have been observed in both simulations and spacecraft data \citep{Ruffolo2006, Perri2007, Perri2008, Effenberger2025}.

In order to interpret the anisotropy observed in our localized study of field line random walk, we compute the in-plane anisotropy and compare it with the minimum variance of the mesoscopic magnetic field
$$
M_{ij} = \langle B_i B_j \rangle_V - \langle B_i \rangle_V\langle B_j \rangle_V, 
\label{eq:mva}
$$
where $\langle \dots \rangle_V$ is an average over an appropriate volume $V$. As typical, we compute the eigenvalues $\{\mu_j\}$, where $\mu_1>>\mu_2>\mu_3$ over a volume $V$ restricted to the near-CME injecting point. Specifically, the mean ratios $\mu_1/\mu_2 \approx 146$ and $\mu_2/\mu_3 \approx 22$ quantitatively demonstrate the presence of strong magnetic anisotropy, being clearly far from unity. 
To roughly establish a link between discussion anisotropy and the magnetic field structure, we compare the anisotropy of the distribution $D^*_1/D^*_2$ with the magnetic field variance at the injection region $\sqrt{\mu_1^2+\mu_2^2+\mu_3^2}$. As shown in Figure~\ref{fig4:diffproperties}-(c), there is a clear correlation. Apparently, the diffusion is strongly affected by the starting region of the field-line tracing. In the case of shock-accelerated gradual SEPs, the field lines of interest are those at the shock, where there is a conspicuous current density. This suggests that the magnetic field is highly sheared at the base. We also compute the correlation by varying the volume $V$, extending it to the entire region covered by the field lines from the CME to 1 AU, confirming the same trend.

To reproduce this observation, we construct a toy model by adding an in-plane sheared field. We integrate the magnetic field line equations in a 3D Cartesian domain, combining a constant background field along the propagation direction with a transverse sheared component that varies linearly across the coordinate axes. To emulate magnetic turbulence, we apply a stochastic random walk to the transverse coordinates at each integration step.	
The model (Figure~\ref{fig4:diffproperties}-(d)) reproduces the anisotropic distribution, with the main axis orthogonal to the gradient direction. This secondary anisotropy, different from the typical parallel-perpendicular anisotropy in solar wind applications, was discussed by \citet{Ruffolo2006} where the anisotropy was due to in-plane correlation lengths of turbulence. In our case, although we use isotropic in-plane correlation lengths, we observe the same effect. This is because shear can be viewed as an external scale that induces correlations \citep{Wan2012}. Such constant shears, although they do not have a dynamical effect here, create a difference in the effective in-plane correlation lengths. This effect is instantaneous in the starting small volume, but field lines keep memory of this effect while they propagate out into the heliosphere.

\section{Discussion and Conclusions}\label{sec:conclusions}
We have presented MICMAC, a novel cross-scale model that combines the mesoscopic EUHFORIA code with a subgrid turbulence model based on non-quasilinear Monte Carlo transport. We have applied this framework to a CME event that misses Earth, demonstrating that cross-scale coupling dramatically alters magnetic connectivity and particle propagation.
	
Turbulence dramatically broadens and distorts the magnetic connection between the CME nose and Earth, producing footpoint distributions that are highly non-Gaussian, anisotropic, and patchy (Figure~\ref{fig3:allcases}). This suggests that the simple ``well-connected'' or ``poorly-connected'' dichotomy used in operational SEP forecasting may be insufficient \citep{Laitinen2018, Whitman2023}. The patchy nature of these distributions is reminiscent of the dropout phenomenon observed in SEP events \citep{Tooprakai2016}, where particles are confined within magnetic flux tubes due to the topological trapping of field lines \citep{Ruffolo2004}.
	
The distribution's enstrophy grows as the CME approaches Earth (Figure~\ref{fig4:diffproperties}-(b)), signaling patchy diffusive behaviour. 
This indicates that the large-scale CME structure imposes a coherent ``memory'' on the turbulent field, preventing the development of fully diffusive transport. Similar memory effects and enhanced cross-field diffusion efficiency in complex solar wind configurations have been reported in \citep{Wijsen2019}, who showed that compression regions can act as magnetic mirrors and facilitate particle trapping.
	
A simple shear-layer toy model reproduces the observed in-plane anisotropy (Figure~\ref{fig4:diffproperties}-(d)), suggesting that local current sheet geometry imprints a persistent memory on the turbulent field. This is consistent with the work of \citet{Ruffolo2006}, who showed that in-plane anisotropy can arise from anisotropic correlation lengths. Here we show that the same effect can be produced by a shear, even with isotropic turbulence modeling. The anisotropy we observe may also be related to the field-line random walk in nonaxisymmetric turbulence, where the diffusion coefficients $D_x$ and $D_y$ are coupled through biquadratic equations \citep{Ruffolo2006}.
	
The present approach is closely related to that adopted by other SEP models that solve the FTEs using stochastic pseudo-particles. We therefore plan to compare directly with existing models, such as PARADISE \citep{wijsen2020paradise}, to better comprehend the specific effects introduced by the cross-scale couplings pof turbulence.


These results advance lunar and heliospheric space weather forecasting, highlighting the necessity of cross-scale CME-turbulence coupling \citep{Whitman2023}. Future work will benchmark our stochastic pseudo-particle approach against codes like PARADISE \citep{wijsen2020paradise} and incorporate deterministic drifts, radial turbulence gradients, and intermittent fields to model energy-dependent transport alongside multi-spacecraft observations.

\section*{Acknowledgments}
Authors acknowledge the Space It Up project funded by the Italian Space Agency, ASI, and the Ministry of University and Research, MUR, under Contract No. 2024-5-E.0 - CUP No. I53D24000060005.
SP is funded by the European Union (ERC-AdG agreement No 101141362, Open SESAME). Views and opinions expressed are, however, those of the author(s) only and do not necessarily reflect those of the European Union or the European Research Council. Neither the European Union nor the granting authority can be held responsible. SP is also funded by the projects C16/24/010 (C1 project Internal Funds KU Leuven), G0B5823N and G002523N (WEAVE) (FWO-Vlaanderen), and 4000145223 (SIDC Data Exploitation (SIDEX2), ESA Prodex).

	\bibliography{bibliografia}{}
	\bibliographystyle{aasjournal}
	
\end{document}